\documentclass[12pt]{article}
\pdfoutput=1 
\usepackage{cmap}

\usepackage[utf8]{inputenc}  
\usepackage[X2,T3,T2A,T1]{fontenc}
\usepackage[russian,ngerman,english]{babel}
\usepackage[dvipsnames]{xcolor} 

\usepackage[square,comma,numbers,compress]{natbib}

\usepackage[pdfpagelabels=false,linktocpage=true,bookmarksnumbered=true,breaklinks=true,pagebackref=false,colorlinks=true,urlcolor=Mahogany,citecolor=Mahogany,linkcolor=Mahogany]{hyperref}

\usepackage{hypernat}

\begin{document}
\bibliographystyle{plainnat}

\vspace{3.5cm}

\renewcommand{\thefootnote}{\fnsymbol{footnote}}
\begin{center}
{\LARGE\baselineskip0.9cm
On selfconsistency in quantum field theory 
\\[1.5cm]}

{\large
K. Scharnhorst\footnote[2]{\begin{minipage}[t]{12cm}
E-mail: {\tt k.scharnhorst@vu.nl},\hfill\ \linebreak
ORCID: \url{http://orcid.org/0000-0003-3355-9663}
\end{minipage}}
}\\[0.3cm]

{\small 
Vrije Universiteit Amsterdam,
Faculty of Sciences, Department of Physics and Astronomy,
De Boelelaan 1081, 1081 HV Amsterdam, The Netherlands}\\[1.5cm]

\begin{abstract}
A bootstrap approach to the effective action in quantum field
theory is discussed which entails the invariance under quantum
fluctuations of the effective action describing physical 
reality (via the S-matrix). 
\end{abstract}

\end{center}

\renewcommand{\thefootnote}{[\alph{footnote}]}
\thispagestyle{empty}

\newpage
\section{\label{introduction}Introduction}

\noindent
These are times of contemplation and reorientation in quantum field theory.
With the experimental detection of the Higgs boson in 2012 finally
the finishing stone of the Standard Model of elementary
particle physics \citep{1992donoghue} surfaced. 
On the theoretical side, the Standard Model
is based on the concept of renormalized local quantum field theory.
The confidence in this concept originally and primarily relies on the
extraordinary success of the centerpiece of the Standard Model, 
quantum electrodynamics (QED), which exhibits an 
impressive agreement between theory and experiment (Cf., e.g.,
\citep{2006mohr}, for more comprehensive reviews see \citep{1990kinoshita}.).
The successful application of renormalized local quantum field theory
to the other components of the Standard Model, the 
electroweak theory and to quantum chromodynamics (QCD), have
further advanced this confidence. On the other hand, many
practicing quantum field theorists are aware of the many 
shortcomings and deficiencies of the concept of renormalized 
local quantum field theory which, by the way, has changed and
developed in a multifold way in the decades since its birth
at the end of the 1940's. To name a few of these issues we
mention here the occurrence of ultraviolet (UV) divergencies, the cosmological
constant problem, hierarchy and naturalness problems, Haag's theorem
(For an instructive illustration of the views of a number of 
well-known quantum field theorist see, e.g., the conference volume
\citep{1999tianyucao}.). It should, however, be pointed out
that in the quantum field theory community there is no unified view
which of these issues constitute features and which are problematic
aspects of renormalized local quantum field theory. Correspondingly,
opinions on which direction should be chosen for the future
conceptual and technical development of quantum field theory 
are diverse (For a recent account of the current situation
see \citep{2022poland}.). While many active researchers might favour new ideas
which have not been discussed in the past a certain fraction
of the quantum field theory community might be willing to not
completely disregard past ideas which have largely been bypassed 
so far. In the present article, it is our intention to bring together
a couple of thoughts and ideas (supplemented by the corresponding references)
that have emerged in the past. We hope that the collection of this
information in a single place will be beneficial to those readers
who consider voices from the past as an inspiration for future 
research rather than purely as a matter for historians of science.\\

Let us start with pointing out that with reference to the UV divergency
problem in QED some of the very fathers of this theory have repeatedly
expressed their dissatisfaction with their own creation up to the end of 
their lives. So, Richard Feynman stated 1965 in his Nobel Prize speech quite 
frankly : ``\protect{\ldots}, I believe there is really no satisfactory 
quantum electrodynamics, but I'm not sure. \ldots, I think that the 
renormalization theory is simply a way to 
sweep the difficulties of the divergences of electrodynamics under the rug.''
(\citep{1966feynman}, {\it Science} p.\ 707, {\it Phys.\ Today} pp.\ 43/44,
{\it Prix Nobel} p.\ 189, {\it Nobel Lectures} p.\ 176, 
{\it Selected Papers} p.\ 30).
Now one might think that Feynman has later, after the development of
the Wilsonian view on renormalization in the early 1970's and the emergence
of the effective field theory concept changed his view. However, 
this seems not to be the case and one can read 
in Feynman's popular science book 
``{\it QED -- The Strange Theory of Light and Matter}''
the passage ($n$ and $j$ are the bare counter parts of the physical electron
mass $m$ and electron charge $e$, respectively.):
``The shell game that we play to find $n$ and $j$ is technically 
called ``renormalization.''
But no matter how clever the word, it is what I would call a dippy process!
Having to resort to such hocus-pocus has prevented us from proving that 
the theory of quantum
electrodynamics is mathematically self-consistent. It's surprising that
the theory still hasn't been proved self-consistent one way or the other
by  now; I suspect that  renormalization  is  not  mathematically
legitimate. What is certain is that we do not have a good
mathematical way to describe the theory of quantum electrodynamics:
such a bunch of words to describe the connection between $n$ and $j$ and
$m$ and $e$ is not good mathematics.''
(\citep{1985feynman}, 1st ed. 1985, pp.\ 128/129, 2nd ed.\ 2006, p.\ 127/128).
In a similar way Paul Dirac stated in a lecture in 1975 (published 1978):
``Hence most physicists are very satisfied with the situation. They say:
``Quantum electrodynamics is a good theory, and we do not have
to worry about it any more.'' I must say that I am very dissatisfied with the
situation, because this so called ``good theory'' does involve 
neglecting infinities which appear in its equations, neglecting them
in an arbitrary way. This is just not sensible mathematics.
Sensible mathematics involves neglecting a quantity when it turns out
to be small -- not neglecting it just because it is infinitely great and
you do not want it.'' (\citep{1978dirac}, p.\ 36).
Few years later, in 1980 (published in 1983), Dirac repeats his
critical view:
``Some new relativistic equations are needed; new kinds of interactions
must be brought into play. When these new equations and new
interactions are thought out, the problems that are now bewildering to
us will get automatically explained, and we should no longer have to
make use of such illogical processes as infinite renormalization. This is
quite nonsense physically, and I have always been opposed to it. It is
just a rule of thumb that gives results. In spite of its successes, one
should be prepared to abandon it completely and look on all the successes
that have been obtained by using the usual forms of quantum
electrodynamics with the infinities removed by artificial processes as
just accidents when they give the right answers, in the same way as the
successes of the Bohr theory are considered merely as accidents when
they turn out to be correct.'' (\citep{1983dirac}, p.\ 55).
The weight one might be tempted to assign to these views certainly 
will depend on the scientific taste of each theoretician, however,
at least one should take note of them.\\

It often happens in the course of the development of science that 
early considerations and ideas are more fundamental than those emerging
later. This is easily explainable by the fact that at the early 
stages of the development of a subject one enters largely 
unchartered territory and simple and structural ideas are then
needed to choose the right road to scientific progress. Sometimes
conflicting ideas are competing with each other. Initial dominance 
of one idea does not necessarily mean that less successful concepts
should be written off. It happens from time to time that these
disregarded concepts make a surprising return for one reason or the other.
Consequently, a look into the past (science history) may be helpful
for shaping the future. For the following considerations, we will depart
from such an element of science history.\\

\section{\label{early}Extending early thoughts of Wolfgang Pauli}

Let us start our concrete discussion with a statement made by Wolfgang
Pauli in a private letter (in German) to Victor Weisskopf (by then,
assistant to Wolfgang Pauli at the ETH Zurich) in 1935. The 
comment of Wolfgang Pauli concerns the self-energy of the electron
in QED, a theory which was under development
in those days. In the context of the struggle with the
UV divergencies of QED Wolfgang Pauli expresses the following conviction
(English translation in brackets: K.\ S.):
``\ldots (Ich glaube allerdings, 
da\ss\ in einer vern\"unftigen Theorie die Selbstenergie
nicht nur endlich, sondern Null sein mu\ss\ \ldots [\ldots (However, 
I believe that in
a sensible theory the self-energy has not only to be finite but zero \ldots]''
(\citep{1993vonmeyenn}, p.\ 779 of: Letter [425b] of December 14, 1935 
from Pauli to Weisskopf. Part of:
Nachtrag zu Band I: 1919–1929 und II: 1930–1939, pp.\ 733-826).
How can we understand this expectation of a future correct 
quantum electrodynamical theory expressed by Wolfgang Pauli?
If one starts quantizing the theory (in this case, charged fermions
interacting with the electromagnetic field) on the basis of a Lagrangian
with the physical (i.e., measured) value of the mass of the fermions
inserted all physical processes that can conceivably have an impact on that
mass have effectively been taken into account already. Consequently,
the total impact of all physical processes taken into account in the
(theoretical) process of quantization (i.e., taking into account
quantum fluctuations) on the mass of these fermions 
should vanish (nonrenormalization). This statement can be 
reformulated by saying that the fermion mass should not receive 
any radiative corrections under quantization. One can now 
extend this early point of view of Wolfgang Pauli
and consider starting quantization not only with
the physical value of the fermion mass in the initial Lagrangian but
choosing as initial Lagrangian (in an arbitrary theory now) 
the (effective) Lagrangian which 
describes the physical world (with all its -- infinitely many -- 
nonlocal and nonpolynomial terms). In principle (in theory, 
not in practice, of course), this can be read off from scattering 
experiments (For the connection between the scattering matrix and
the effective action see, e.g., \citep{1978slavnov}, sec.\ 2.4, 
\citep{1988jevicki}.). 
If one now starts the process of quantization with this
``true'' Lagrangian all radiative corrections to it should vanish
because any quantum fluctuations described by this Lagrangian have
already been taken into account in this Lagrangian.
Consequently, the physical (effective)
Lagrangian should be invariant under the process of quantization,
all radiative corrections should vanish. 
This view of the process of quantization amounts to bootstrapping the  
effective action of a theory. Quantities denoted within
standard local renormalizable quantum field theory as
bare and renormalized ones, respectively, then coincide.\\

Before continuing our verbal discussion, let us make the above 
statements more precise in terms of equations.
We consider within a path integral framework
Lagrangian quantum field theory in flat ($n$-dimensional Minkowski) space-time 
and a (one-compo\-nent) scalar field theory to pursue the discussion
(for the following equations cf., e.g., \citep{1980itzykson}, chap.\ 9).
A generalization to more complicated theories is straightforward.
The generating functional of Green
functions of the scalar field $\phi(x)$ is given by the equation
\parindent0.cm

\begin{equation}
\label{BA1}
Z[J] =\ C \int D\phi\ \ {\rm e}^{\displaystyle\ 
i\Gamma_0 [\phi]\ +\ i \int d^nx\ J(x) \phi (x)}
\hspace{1.5cm},\hspace{0.5cm}\\
\end{equation}

where $\Gamma_0 [\phi]$ is the so-called classical action of
the theory and $C$ some fixed normalization constant. 
Then, the generating functional of the connected Green functions is

\begin{equation}
\label{BA2}
W[J]\ =\ -i \ln Z[J] \hspace{1.5cm}.\hspace{0.5cm} \\
\end{equation}

The effective action $\Gamma [\bar\phi]$, which also is
the generating functional of the one-particle-irreducible
(1PI) Green functions, is obtained as the first Legendre
transform of $W[J]$,

\begin{equation}
\label{BA3}
\Gamma [\bar\phi]\ =\ W[J] - \int d^nx\ J(x) \bar\phi (x)
\ \ \ \ \ . \\
\end{equation}

Here

\begin{equation}
\bar\phi (x)\ =\ {\delta W[J]\over \delta J(x)} \\
\end{equation}

which in turn leads to

\begin{equation}
\label{BA4}
{\delta \Gamma [\bar\phi]\over \delta \bar\phi (x)}\ =\ -\ J(x)\\
\end{equation}

in analogy to the classical field equation for $\Gamma_0 [\phi]$.
Equivalently, using the above expressions 

\begin{equation}
\label{BA5}
{\rm e}^{\displaystyle\ i \Gamma [\bar\phi]}\ \ =\ C
\int D\phi\ \ {\rm e}^{\displaystyle\ i\Gamma_0 
[\phi + \bar\phi]\ +\ i \int d^nx\ J(x) \phi (x)}\\
\end{equation}

can be considered as the defining relation for the effective action,
where the r.h.s.\ of the above equation has to be
calculated using a current $J(x)$, given by Eq.\ (\ref{BA4}),
which is a functional of $\bar\phi $.
Eq.\ (\ref{BA1}) defines a map,
$g_1: \Gamma_0 [\phi ]\longrightarrow Z[J]$, from the 
class of functionals called classical actions to the 
class of functionals $Z$. Furthermore, we have 
mappings, $g_2: Z[J]\longrightarrow W[J]$, (Eq.\ (\ref{BA2})) and 
$g_3: W[J]\longrightarrow \Gamma [\bar\phi]$
(Eq.\ (\ref{BA3})). These three maps together define a
map $g_3\circ g_2\circ g_1 = f: \Gamma_0 [\phi ]
\longrightarrow \Gamma [\bar\phi] $ (Eq.\ (\ref{BA5}))
from the set of so-called classical actions to the set of 
effective actions. It is understood that in order to properly
define the map a regularization scheme for the scalar field
theory is applied. Up to renormalization effects,
the classical action $\Gamma_0 [\phi]$ determines the effective
action $\Gamma [\bar{\phi}]$ uniquely via the map $f$ which encodes 
quantum principles. In this standard scheme, 
the (quantum) effective action is built 
directly from classical physics and exhibits no independence in
its own right.\\

The terms of the difference 
$\Delta \Gamma [\bar\phi] =\Gamma [\bar\phi]\ -\ \Gamma_0 [\bar\phi]$ are
denoted as 'radiative corrections'. The above verbal reasoning
in generalization of early thoughts of Wolfgang Pauli leads to the equation  

\begin{equation}
\Delta \Gamma [\bar\phi] = 0\ ,
\end{equation}

expressing the vanishing of all radiative corrections, i.e.,

\begin{equation}
\Gamma [\bar\phi]\ =\ \Gamma_0 [\bar\phi]\ .
\end{equation}

The equation for the complete effective action which
is equivalent to the fixed point condition for the
map $f$ reads ($C^\prime$ is some normalization constant)

\begin{equation}
\label{BB6}
{\rm e}^{\displaystyle\ i \Gamma [\bar\phi]}\ \ =\ C^\prime\
\int D\phi\ \ {\rm e}^{\displaystyle\ i\Gamma [\phi + \bar\phi]\
+\ i \int d^nx\ J(x) \phi (x)}
\hspace{1.5cm} ,\hspace{0.5cm}\\
\end{equation}

\parindent0.cm
where

\begin{equation}
\label{BB7}
J(x) \ =\ -\ {\delta \Gamma [\bar\phi]\over \delta \bar\phi (x)}
\hspace{1.5cm} .\hspace{0.5cm}\\
\end{equation}
 
The above selfconsistency equation (\ref{BB6}) defines the 
(finite) effective action
(including its coupling constants and mass ratios)
without any reference to classical physics 
exclusively via quantum principles encoded in the map $f$.
The fixed points of the map $f$ then describe physical reality.
From out this perspective, the standard formulation of quantum field 
theory represented by eq.\ (\ref{BA1}) can roughly be understood as the 
first step of an iterative solution of the nonlinear
functional integro-differential equation (\ref{BB6}) by applying
the map $f$ to some initial (in this case `classical') action $\Gamma_0 [\phi]$.
For the first time, the above equation (\ref{BB6})
to be taken as basis of quantum field theory 
has been proposed in 1972 by L.\ V.\ Prokhorov \citep{1972prokhorov}. 
Not being aware of the earlier work by Prokhorov, the same proposal has 
been made by the present author in 1993 \citep{1997scharnhorst}.
In a somewhat different (Hamiltonian) setting (coupled cluster
methods), J.\ S.\ Arponen has
expressed similar ideas in 1990 (\citep{1991arponen}, p.\ 173, 
paragraph starting with the words: ``The possible solution corresponds 
to a system which suffers no change under quantization.'').\\

\section{\label{disc}Further discussion}

Given the mature state of standard renormalizable quantum field
theory, the above point of view (defining the effective action as 
a fixed point of the map $f$) faces myriads of objections.
Some of them may be misunderstandings, others are completely 
justified concerns, others yet are possibly prejudices. Misunderstandings
can be dealt with most easily -- by clarifications. For example,
one might ask: Given the crucial role of radiative corrections
within the standard formulation of quantum field theory in correctly
describing physical reality (for example, in QED) how could one 
ever possibly think of a theory of physical reality characterized by the 
vanishing of all radiative corrections (in an interacting theory)? 
The difficulty here, however, is 
just a terminological one. Of course, also a modified formulation of 
quantum field theory needs to deliver those kind of terms in the
effective action we denote as radiative corrections within the
established standard approach. While the
analytical expression yielded from a modified formulation of quantum
field theory may differ from those within the standard formulation, 
the numerical results for experimentally accessible quantities (e.g., the 
anomalous magnetic moment of the electron) need to be (almost --
within experimental limits) the same. The point is that in the 
modified view of quantum field theory represented by eq.\ (\ref{BB6})
those terms denoted in the standard formulation as radiative corrections
are already incorporated in the action to be quantized. But, as the 
action to be quantized is supposed to be invariant under quantization
(according to eq.\ (\ref{BB6})) no new terms may emerge, consequently,
in the modified formulation of quantum field theory no radiative 
corrections (relative to the initial action to be quantized) occur.\\

Certainly, one elementary and justified concern with respect to eq.\ (\ref{BB6})
consists in the question of whether eq.\ (\ref{BB6}) allows any non-trivial
(i.e., non-Gaussian) solutions (Free field theories, of course, always
obey eq.\ (\ref{BB6}).). In fact, it has been shown by example 
in a finite-dimensional Grassmann algebra
analogue of eq.\ (\ref{BB6}) (i.e., within a fermionic zero-dimensional field
theory) that eq.\ (\ref{BB6}) has exact non-trivial (i.e., non-Gaussian)
solutions \citep{2003scharnhorst}. For an (bosonic) example from
standard analysis see \citep{2003pioline}. Of course, as has been pointed
out by Prokhorov \citep{1972prokhorov} from the outset 
eq.\ (\ref{BB6}) represents a very complicated equation and presently
very little can be said about its eventual non-trivial solutions in general.
Experience from effective action studies in quantum field theory 
tells us that non-trivial solutions of eq.\ (\ref{BB6}) can be
expected to be nonlocal and nonpolynomial functionals of fields.
Whether these nonlocal and nonpolynomial actions $\Gamma$ 
solving eq.\ (\ref{BB6})
preserve unitarity and causality can only be decided once they are
found. However, it has been shown for a wide class of
nonlocal and nonpolynomial (scalar) quantum field theories in the past
\citep{1973alebastrov,1974alebastrov} that they respect these two 
principles -- a fact, from which one may derive certain optimism.
It is of course conceivable that the general version of eq.\ (\ref{BB6})
(written down for an arbitrary but fixed collection of fluctuating fields) does 
not have any non-Gaussian solution at all for the certain field content 
one has chosen. If this was the case the existence of a non-Gaussian 
solution to the generalised version of eq.\ (\ref{BB6}) could be
applied as a theory selection criterion perhaps in the
same way as the (stationary) Schr\"odinger equation selects 
energy (eigen)values of quantum mechanical systems. In a certain 
sense, at the end of 
the day only non-Gaussian solutions of eq.\ (\ref{BB6}) are physical ones
because only they provide the interactions for the structures we 
observe in physical reality. Away from the rigid concept discussed
above of the effective action as a fixed point of the map $f$ non-Gaussian
solutions of eq.\ (\ref{BB6}) may be considered also interesting within the
standard lore. While usually perturbation theory is built around a
Gaussian solution of eq.\ (\ref{BB6}), choosing a non-Gaussian solution
of eq.\ (\ref{BB6}) as starting point for perturbation theory may also be
of some interest. For a discussion in this direction see \citep{2018kuehn}.\\ 

From a methodological point of view, the largest difference of the
approach represented by eq.\ (\ref{BB6}) to the established approach 
in standard quantum field theory consists in 
the following. Standard local renormalizable quantum field theory 
starts (among other things, e.g.,
choosing the space-time dimensionality) with the choice of the field content 
of the theory under consideration and the functional form of the 
(classical/bare) action $\Gamma_0$ (cf.\ eq.\ (\ref{BA1})), 
a quantity which, in principle, is unobservable (due to the 
existence of radiative corrections). This is also true
in the different versions of the 
Wilsonian approach to quantum field theory (inspired by the 
theory of phase transitions in statistical mechanics). While in
a statistical mechanical system (e.g., a spin system modelling
a certain microscopic condensed matter structure) the structure of
the Hamiltonian defined on a lattice with fixed lattice spacing 
can in principle be linked to experimental data, this is not the case
within quantum field theory where the bare action is
not related to observation
(For a related discussion see, e.g., \citep{2019rosaler}.). 
Consequently, in the established standard approach to quantum field
theory the theoretical description of physical reality
(i.e., the effective action $\Gamma$) is inferred from 
quantities not accessible to experiment in principle. 
In contrast, within the approach represented by eq.\ (\ref{BB6})
only the field content of the quantum fluctuations can be chosen,
the functional form of the effective action $\Gamma$ is selfconsistently
restricted by its property to be a solution of this equation. Beyond free
field theories, i.e., for non-Gaussian solutions of eq.\ (\ref{BB6}),
this can be expected to be highly restrictive.\\

\section*{Acknowledgement\phantomsection}
\addcontentsline{toc}{section}{Acknowledgement}

Kind hospitality at 
the Department of Physics and Astronomy of the Vrije Universiteit 
Amsterdam is gratefully acknowledged.\\

\pagebreak
\section*{\phantomsection}

\end{document}